\RequirePackage{fix-cm}
\RequirePackage{amsmath}
\documentclass[twocolumn,epjc3]{svjour3}  
\smartqed  
\RequirePackage{graphicx}
\RequirePackage{mathptmx}      
 
\usepackage{color}
\usepackage{mathrsfs}
\usepackage{amssymb, bm}

\newcommand{\be}{\begin{equation}}
\newcommand{\ee}{\end{equation}}

\begin{document}

\title{Helmoltz problem for the Riccati equation from an analogous 
Friedmann equation  
}


\author{Valerio Faraoni \thanksref{e1,addr1}
}


\thankstext{e1}{e-mail: vfaraoni@ubishops.ca}


\institute{Department of Physics \& Astronomy, Bishop's University, 
2600 College Street, Sherbrooke, Qu\'ebec, Canada J1M~1Z7 \label{addr1}
}

\date{Received: date / Accepted: date}

\maketitle

\begin{abstract} 

We report a solution of the inverse Lagrangian problem for the first order 
Riccati differential equation by means of an analogy with the Friedmann 
equation of a suitable Friedmann-Lema\^itre-Robertson-Walker universe 
in 
general relativity. This analogous universe has fine-tuned parameters and 
is unphysical, but it suggests a Lagrangian and a Hamiltonian for the 
Riccati equation and for the many physical systems described by it.


\end{abstract}

\section{Introduction}
\label{sec:1}
\setcounter{equation}{0}

The inverse variational problem, or Helmoltz problem, for a set of 
ordinary differential equations consists of finding a Lagrangian such that 
the associated Euler--Lagrange equations reproduce the given system.  
Although necessary and sufficient conditions for solving the inverse 
variational problem were already given by Helmoltz in 1887 \cite{Helmoltz} 
they are rather involved \cite{Douglas,Sarlet} and, in general, 
determining whether they are satisfied and finding the explicit solution 
is cumbersome and may require one to solve a large system of equations 
({\em e.g.}, \cite{Prince,Zenkov}).

An alternative, unconventional, approach to the Helmoltz problem is 
provided by analogies: if a certain differential equation is analogous to 
another one ruling an analogous system which admits a known Lagrangian or 
Hamiltonian formulation, the association of a Lagrangian with the original 
equation is straightforward. Of course, this approach can only be applied 
to systems analogous to other systems with known Lagrangian or Hamiltonian 
formulations, that is, in exceptional cases. Here we consider the first 
order non-linear Riccati equation \cite{Ince,Hille}
\be
\frac{du(x)}{dx} + c_0 \, u^2(x) +c_1=0 \,, \label{Riccati}
\ee
where $c_{0,1}$ are constants, which describes many physical systems 
including, {\em e.g.}, falling raindrops \cite{Rosu,myexbook} or charges 
in constant electric field, avalanches \cite{Pudasaini}, debris slides 
\cite{Pudasaini2}, geomagnetic fields \cite{geomagnetic}, and box models 
of ocean basins \cite{DenmarkStrait}; the Riccati equation is also related 
to the Schr\"odinger, the Ermakov-Pinney, and other equations of 
fundamental physics \cite{Schuch,Schuch2}. A second order equation is 
naturally associated with Eq.~(\ref{Riccati}): its generalizations, 
including 
systems of non-linear oscillators, and their Lagrangian formulations have 
been studied, also in relation with integrability and 
superintegrability \cite{Carinena,Musielak,Cieslinski,Musielak2}. Here, 
in a different context, we solve the Helmoltz problem of 
Eq.~(\ref{Riccati}) by means of an analogous Friedmann equation. The 
latter describes a suitable universe in spatially homogeneous and 
isotropic (or Friedmann-Lema\^itre-Robertson-Walker, in short FLRW) 
cosmology, for which a Lagrangian is known. In order for the analogy to 
hold, one must impose a fine-tuned relation between the parameters of the 
analogous cosmos (equation of state parameter, energy density of the 
cosmic fluid, cosmological constant, and curvature index), hence the 
analogous universe is not physically relevant {\em per se}. However, this 
is not an issue here since we are not attempting to describe the real 
universe, but we are interested in solving the inverse Lagrangian problem 
for the Riccati equation~(\ref{Riccati}).

The next section discusses in detail the analogy between Riccati and 
Friedmann equations; Sec.~\ref{sec:3} exploits this analogy to solve the 
Helmoltz problem for the Riccati equation, while Sec.~\ref{sec:4} contains 
some concluding remarks. We adopt the notation of Ref.~\cite{Wald}: the 
metric signature is ${-}{+}{+}{+}$, $G$ is Newton's constant, and units 
are used in which the speed of light is unity.

\section{A cosmological analogy for the Riccati equation}
\label{sec:2}

Spatially homogeneous and isotropic cosmologies are described by the FLRW 
line element 
\be
ds^2 =-dt^2 + a^2(t) \left(  \frac{dr^2}{1-Kr^2} +r^2 d\Omega_{(2)}^2 
\right) 
\ee
in comoving polar coordinates $\left( t,r,\vartheta, \varphi \right) $, 
where $d\Omega_{(2)}^2 = d\vartheta^2 + \sin^2 \vartheta \, d\varphi^2$ is 
the line element on the unit 2-sphere, $K$ is the curvature index  
normalized to $0, \pm 1$, and 
$a(t)$ is the scale factor describing the expansion history of the 
universe \cite{Wald,Liddle}. We assume that the latter is filled 
with a perfect fluid with energy density $\rho(t)$ and isotropic pressure 
$P(t)$ related by the barotropic, linear, and constant equation of state
\be 
P=w\rho \,, \quad \quad w=\mbox{const.}   \label{eos}
\ee
The evolution of the scale factor $a(t)$ and of $\rho(t)$ and $P(t)$ is 
ruled by the  Einstein-Friedmann equations \cite{Wald,Liddle}
\be
\left( \frac{ \dot{a} }{a} \right)^2 = \frac{8 \pi  G}{3} \, \rho 
-\frac{K}{a^2} +\frac{\Lambda}{3} \,,\label{Friedmann}
\ee

\be
\frac{\ddot{a}}{a}  =-\frac{4\pi G}{3} \left( \rho +3P\right) 
+\frac{\Lambda}{3} \,,\label{acceleration}
\ee

\be
\dot{\rho} +3H \left( P+\rho \right)=0 \,,\label{conservation}
\ee
where an overdot denotes differentiation with respect to the cosmic (or 
``comoving'') time $t$ and $\Lambda$ is the cosmological constant. 

It is well-known \cite{Faraoni:1999qu,Barrow1993,JantzenUggla} that  
by combining the 
Friedmann equation~(\ref{Friedmann}) and the acceleration 
equation~(\ref{acceleration})   written in terms of the 
conformal time $\eta$ (defined 
by $dt\equiv a d\eta$), one obtains a Riccati 
equation~(\ref{Riccati}) 
(a similar coordinate transformation has been 
known for the two-body problem since the times of Euler 
\cite{Euler,Bohlin1911,Sundman1912,LeviCivita1920,Bond1985,HeggieHut2003}).  
Here we pose instead the question of whether the 
Friedmann equation {\em in cosmic time} can assume 
the Riccati 
form~(\ref{Riccati}). The answer is affirmative, but this only happens 
when the fluid has (phantom) equation of state parameter $w=-5/3$, 
hyperbolic spatial sections with  $K=-1$, 
and (fine-tuned) cosmological constant $\Lambda\neq 0$, or 
when $w=1/3$, $K=-1$, and 
$\Lambda>0$. We derive this result in the 
following.

Assuming the equation of state~(\ref{eos}), the covariant 
conservation equation~(\ref{conservation}) is integrated to 
\cite{Wald,Liddle} 
\be
\rho(a) = \frac{\rho_0}{ a^{3(w+1)}} \,,
\ee
where $\rho_0>0$ is a constant. Then the Friedmann 
equation~(\ref{Friedmann}) is recast as 
\be
\dot{a} = \pm \sqrt{ \frac{8\pi G}{3} \, \rho_0 a^{-(3w+1)} - K 
+\frac{\Lambda}{3} \, a^2} \,,  \label{X}
\ee
where the argument of the square root is necessarily non-negative if the  
Friedmann equation~(\ref{Friedmann}) is to admit solutions. We now ask 
when this argument is a perfect square: 
there are three possibilities for this to happen. The 
first case 
corresponds to $K=-1$ and $w=-5/3$ and allows 
one to write the argument of the square root as 
\be
\left( \sqrt{ \frac{8\pi G \rho_0}{3} } a^2  \right)^2 
+ (\sqrt{1})^2 
+\frac{\Lambda}{3}\,  a^2 \,;
\ee
then we set $\frac{ \Lambda a^2}{3} = \pm 2 \sqrt{1} \sqrt{\frac{8\pi G 
\rho_0}{3} } \, a^2$, which is satisfied only by tuning the cosmological 
constant to one of the two values 
\be
\Lambda =\pm 6 \sqrt{\frac{8\pi G \rho_0}{3}} 
\ee
and then
\be
\dot{a} = \pm \left( \sqrt{ \frac{8\pi G \rho_0}{3}} \, a^2  \pm 1 \right) 
\,,
\ee 
where the two $\pm$ signs are independent, {\em i.e.}, there are four 
possible solutions here.

The second possibility appears for $K=-1$, $w=1/3$, and $\Lambda>0$. In 
this case we identify
\begin{eqnarray}
\frac{8\pi G \rho_0}{3 a^2} + 1 +\frac{\Lambda}{3} \, a^2 &=& \left(
\sqrt{ \frac{8\pi G \rho_0}{3} } \, \frac{1}{a} \right)^2 + \left( \sqrt{ 
\frac{\Lambda}{3} } \, a \right)^2 \nonumber\\
&&\nonumber\\
&\, & + 2 \sqrt{ \frac{8\pi 
G\rho_0 \Lambda}{9}} \nonumber\\
&&\nonumber\\
& =& \left( \sqrt{ \frac{8\pi G \rho_0}{3} } \, \frac{1}{a}  
+\sqrt{\frac{\Lambda}{3} } \, a \right)^2
\end{eqnarray}
provided that 
\be
 \rho_0 \Lambda = \frac{9}{32\pi G} \,,
\ee
and then 
\be
\dot{a} = \pm \left( \sqrt{ \frac{8\pi G \rho_0}{3}} \, \frac{1}{a}
+\frac{1}{2} \sqrt{ \frac{3}{8\pi G \rho_0} } \, a \right)\,.
\ee

The third possibility occurs for $K=+1$, $w=1/3$, and  $\Lambda 
=\frac{9}{32\pi G \rho_0}  >0$, which yields the identification
\begin{eqnarray}
\frac{8\pi G \rho_0}{3 a^2} - 1 +\frac{\Lambda}{3} \, a^2 &=& \left(
\sqrt{ \frac{8\pi G \rho_0}{3} } \, \frac{1}{a} \right)^2 + \left( \sqrt{ 
\frac{\Lambda}{3} } \, a \right)^2 \nonumber\\
&&\nonumber\\
&\, & - 2 \sqrt{ \frac{8\pi 
G\rho_0 \Lambda}{9}} \nonumber\\
&&\nonumber\\
& =& \left( \sqrt{ \frac{8\pi G \rho_0}{3} } \, \frac{1}{a}  
-\sqrt{\frac{\Lambda}{3} } \, a \right)^2 \,,
\end{eqnarray}
therefore,
\be
\dot{a} = \pm \left( \sqrt{ \frac{8\pi G \rho_0}{3} } \, \frac{1}{a} - 
\frac{1}{2} \, \sqrt{ \frac{3}{8\pi G \rho_0} } \,  a \right) \,.
\ee

These fine-tunings between $\Lambda$ and 
the initial  condition  on the energy density 
($\rho_0$)  are  
clearly unphysical and the discussion is 
purely of mathematical interest unless some physical mechanism is found 
that achieves the tuning, which seems unlikely. 

The differential equations for the scale factor can be solved 
directly by quadratures, but here we reduce them to Riccati equations 
because our goal is to solve the Helmoltz problem for the Riccati 
equation.

\subsection{$\Lambda = \pm 4 \sqrt{6\pi G\rho_0}$  } 

In the first case, we have the equation 
\be
\dot{a} = \pm \left( \sqrt{ \frac{8\pi G}{3} \, \rho_0 } \, a^2 \pm 1 
\right) \,,
\ee  
where the two $\pm$ signs are independent, {\em i.e.}, there are four 
equations. Consider first the two possibilities resulting from the 
equation
\be
\dot{a} = \pm \sqrt{ \frac{8\pi G}{3} \, \rho_0 } \, a^2 + 1  \,,
\ee  
which is of the Riccati form~(\ref{Riccati}) with
\begin{eqnarray}
c_0 &=& \mp \sqrt{\frac{8\pi G}{3} \, \rho_0} \,,\\
&&\nonumber\\
 c_1  & = & -1 \,. 
\end{eqnarray}
This Riccati equation is solved by setting \cite{Ince,Hille}
\be
a(t) \equiv \frac{1}{c_0} \, \frac{ \dot{v}}{v} \,,
\ee
which yields
\be
\ddot{v} + c_0 \, c_1  v = 0 \,. \label{eq1forv}
\ee
We discuss separately the two possibilities corresponding to upper and 
lower sign for $c_0$.

\subsubsection{Upper sign}

By choosing the upper sign we have a positive (and fine-tuned) 
cosmological constant $\Lambda$, $c_0 \, c_1>0$, and the solution of the 
resulting harmonic oscillator equation~(\ref{eq1forv}) is $ v(t)= A 
\sin\left( \sqrt{c_0 \, c_1} \, t \right) + B \cos\left( \sqrt{c_0 \, c_1} 
\, t \right)$ where $A$ and $B$ are integration constants, yielding
\be
a(t) =\sqrt{ \frac{c_1}{c_0} } \, \frac{ B \sin\left( \sqrt{c_0 \, c_1} \, 
t \right) -A \cos\left( \sqrt{c_0 \, c_1} \, t \right) }{
A\sin\left( \sqrt{c_0 \, c_1} \, t \right)
+B \cos\left( \sqrt{c_0 \, c_1} \, t \right) } \,,\label{general1}
\ee
where $a(t)$ must be non-negative. Although there 
are two arbitrary integration constants $A$ and $B$ for 
Eq.~(\ref{eq1forv}), in practice there is  only one 
arbitrary initial condition $A/B$ or $B/A$, corresponding to the fact that 
the equivalent Riccati equation  is of first order. Special initial 
conditions give particular solutions.

\noindent $\bullet$ $A=0$, $B\neq 0$

In this case the solution~(\ref{general1}) becomes
\be
a(t) = \left( \frac{3}{8\pi G \rho_0} \right)^{1/4} \tan \left[ \left( 
\frac{8\pi G \rho_0}{3} \right)^{1/4} \, t \right]
\ee  
in the range 
\be
0\leq t <\frac{\pi}{2} \left( \frac{3}{8\pi G \rho_0} \right)^{1/4} 
\equiv t_* \,,
\ee   
which represents a universe starting at a Big Bang and ending 
in a Big Rip ($a\rightarrow +\infty$) at a finite future $t_*$.  
Here we denote as ``Big Bang'' a zero of the scale factor $a(t) 
\rightarrow 0$, but this word does not have the usual textbook meaning in 
the sense 
that the derivative $\dot{a}$, the energy density $\rho$, and the pressure 
$P$ do not diverge at this 
``Big Bang''. Here the singularity is ``soft'' in the sense that 
$\dot{a}(0)$ is 
finite (but the Hubble 
function $H(t)\equiv \dot{a}/a$ still diverges). Likewise, the Big 
Rip singularity  is not an inverse power-law, as it would happen if only 
the phantom fluid were present \cite{Caldwell:1999ew,Caldwell:2003vq}, but 
has an unusual tangent-like divergence.

\noindent $\bullet$ $A\neq 0$, $B=0$

The solution~(\ref{general1}) becomes 
\be
a(t) = - \left( \frac{3}{8\pi G \rho_0} \right)^{1/4} \cot \left[ 
\left( \frac{8\pi G \rho_0}{3} \right)^{1/4} \, t \right]
\ee  
in the range 
\be
 t_* <t < 2 t_* \,;
\ee
there are again a Big Bang (in the sense that  $a=0$ 
at $t=t_*$)  and a Big Rip where 
$a\rightarrow +\infty$ as $t\rightarrow 2t_*$.

\noindent $\bullet$ $A=B \neq 0$

In this case the solution~(\ref{general1}) takes the form
\be
a(t) = - \left( \frac{3}{8\pi G \rho_0} \right)^{1/4} 
\frac{ \cos
\left[ 2\left( \frac{8\pi G \rho_0}{3} \right)^{1/4} \, t \right]}{
1+\sin \left[ 2\left( \frac{8\pi G \rho_0}{3} \right)^{1/4} \, t 
\right] } 
\ee 
in the range 
\be
 \frac{t_*}{2} <t < \frac{3 t_*}{2} \,.
\ee
Again, the solution begins in a ``soft'' Big Bang (in the sense 
that $a(t)\rightarrow 0$) and ends in a Big Rip 
singularity $a\rightarrow +\infty$.

\subsubsection{Lower sign}

In this case $\Lambda<0$, $c_0>0$, $c_1<0$, and 
$ v(t) = A\mbox{e}^{ \sqrt{c_0|c_1|} \, t} +B \mbox{e}^{ -\sqrt{c_0|c_1|} 
\, t} $ (with $A,B$ integration constants), yielding
\be
a(t) = \sqrt{ \frac{|c_1|}{c_0} } \, \frac{ 
A\mbox{e}^{ \sqrt{c_0|c_1|} \, t} -B \mbox{e}^{ -\sqrt{c_0|c_1|} 
\, t}  }{
A\mbox{e}^{ \sqrt{c_0|c_1|} \, t} +B \mbox{e}^{ -\sqrt{c_0|c_1|} 
\, t} }\,. \label{generallower}
\ee
Contrary to the previous situation (upper sign in Eq.~(\ref{eq1forv})), 
for $A= 0$, $B\neq 0$ the scale factor is negative, therefore we discard 
this possibility and we assume that $A\neq 0$. Special 
initial conditions include the following.

\noindent $\bullet$  $A\neq 0$, $B=0$

In this case the solution is the static universe with 
\be
a(t) = \sqrt{ \frac{|c_1|}{c_0} } =
\left( \frac{3}{8\pi G \rho_0} \right)^{1/4} \equiv a_*  \label{static}
\ee  
resulting from the balance of the negative cosmological constant 
with the 
repulsive phantom fluid and the curvature term in the Friedmann 
equation~(\ref{Friedmann}). The general solution~(\ref{generallower}) with 
$A\neq 0$ asymptotes to $a_*$ at late times $t\rightarrow +\infty$, 
irrespective of the value of this integration constant, therefore the 
solution $a(t) \equiv a_* $ is stable with respect to homogeneous 
perturbations and is a late-time attractor in the phase space of 
the solutions.

\noindent $\bullet$  $A=B \neq 0$

With this choice of initial conditions, the scale factor is
\begin{eqnarray}
a(t) &=& \sqrt{ \frac{|c_1|}{c_0} } \, \tanh \left( \sqrt{|c_1|c_0} \, t 
\right) \nonumber\\
&&\nonumber\\
&=& a_* \,  \tanh \left[ \left( \frac{8\pi G \rho_0}{3} 
\right)^{1/4} \, t \right]   \,. 
\end{eqnarray}
This universe begins from a Big Bang (again, in the 
sense $a\rightarrow 0$) at $t=0$ and evolves for an 
infinite 
time, with the scale factor asymptoting to $a_*$ as $t\rightarrow + 
\infty$. In this case $a(t)$ is analogous to the speed of a raindrop 
falling vertically from rest in a constant gravitational field and 
reaching terminal speed \cite{Rosu,myexbook}, as $a(t) \simeq a_*$ when 
$t\rightarrow +\infty$.

\noindent $\bullet$ $A=-B \neq 0$

In this case, the scale factor 
\be
a(t) = \left( \frac{3}{8\pi G \rho_0} \right)^{1/4} 
\coth  \left[ \left( \frac{8\pi G \rho_0}{3} \right)^{1/4} \, t 
\right] 
\ee  
corresponds to the unusual contracting branch of a pole of the 
scale factor $a(t)$, where the universe begins from infinite size at 
$t=0$ and decreases monotonically, asymptoting to the constant value 
$a_*$ as $t\rightarrow +\infty$. Again, the static universe~(\ref{static}) 
is a late-time attractor in phase space.

\subsubsection{The remaining sign possibilities} 

In this case  we have 
\be
\dot{a} = \pm \left( \sqrt{ \frac{8\pi G\rho_0}{3}}  \, a^2 -1 \right) 
\,,
\ee
which has the solution
\be
a(t) = \pm \left( \frac{3}{8\pi G \rho_0} \right)^{1/4} \arctan \left[
\left( \frac{8\pi G \rho_0}{3} \right)^{1/4} \, (t-t_0)\right] 
\ee
where the range of $t$ is chosen so that $a(t)$ is non-negative.

\subsection{$\rho_0 \Lambda = 9/(32\pi G)$}

We have two possibilities corresponding to this fine-tuned choice of 
$\rho_0$. In the first case,  Eq.~(\ref{X}) assumes the form
\be
\dot{a} = \pm \left( \frac{ \sqrt{A}}{a} + \frac{a}{2\sqrt{A}} \right)
 \,,
\ee
where $A\equiv \sqrt{ 8\pi G \rho_0/3}$. This is  is not  a Riccati 
equation and is readily integrated, giving the solution 
\be
a_{( \pm )} (t) =  \sqrt{ \mbox{e}^{ \pm \, \frac{(t-t_0)}{\sqrt{A} }} 
-2A } 
\,,
\ee
where $t_0$ is an integration constant. 

In the second case,  Eq.~(\ref{X}) becomes
\be
\dot{a} = \pm \left( \frac{ \sqrt{A}}{a} - \frac{a}{2\sqrt{A}} \right)
 \,,
\ee
which admits the solutions
\be
a_{( \pm )} (t) =  \sqrt{ 2A - \mbox{e}^{ \mp \,\frac{(t-t_0)}{\sqrt{A}} }  
} \,.
\ee

\section{Solving the inverse Lagrangian problem for the Riccati equation}
\label{sec:3}

We only need one of the cosmological solutions to find a Lagrangian 
for the Riccati equation~(\ref{Riccati}). Based on the cosmological 
analogy illustrated in the previous section,  focus on the case in 
which the Riccati equation~(\ref{Riccati}) is the same as the Friedmann 
equation for  a FLRW universe permeated by a perfect phantom fluid with 
$P=-5\rho/3$, 
$K= -1 =  -c_1^2$, and $\Lambda= 6 \,c_0 \, c_1$. This cosmic 
analogy inspires an 
unconventional solution of the inverse Lagrangian (or Helmoltz) problem of 
finding a Lagrangian and a Hamiltonian for the Riccati 
equation~(\ref{Riccati}). The standard Lagrangian for FLRW 
cosmology\footnote{The Lagrangian can also be obtained by using the 
lapse function (one obtains the Friedmann equation by varying the action 
with respect to the lapse function).} 
\cite{deRitis1,deRitis2,deRitis3,SanyalModak,mybook} \be 
L \left( a, \dot{a} \right) =  a\dot{a}^2 +\frac{8\pi G}{3} \, a^3 \rho 
-Ka + \frac{\Lambda}{3} \, a^2
\ee
suggests to use 
\be
L_1 \left( u, \dot{u} \right) = u\dot{u}^2 +c_0^2 u^5 +c_1^2 u + 2 \,c_0 
\, c_1  u^2  \label{L1}
\ee
as a Lagrangian for the Riccati equation. The corresponding Hamiltonian is
\be
{\cal H}_1 = \dot{u} \, \frac{\partial L_1}{\partial \dot{u}} -L_1= 
u\dot{u}^2 -c_0^2 u^5 -c_1^2 u -2 \, c_0 \, c_1 
u^2 \,.
\ee
Since this Hamiltonian does not depend explicitly on time it is conserved, 
yielding the Beltrami identity ${\cal H}_1=$~const. To 
actually reproduce the Riccati equation, one must choose this constant to 
be zero, according to the fact that the dynamics of general relativity is 
constrained \cite{Wald}. In cosmology, this fact is reflected in the fact 
that the 
Friedmann equation is a first order constraint, not  a full (second order) 
equation of motion, and the vanishing of ${\cal H}_1$ enforces precisely 
this constraint (``Hamiltonian constraint'') \cite{Wald,Liddle}. Setting
${\cal H}_1=0 $ yields 
\be
\dot{u}= \pm \left( c_0 \, u^2 +c_1 \right) \,.
\ee
Choosing the lower sign reproduces the Riccati equation~(\ref{Riccati}), 
while choosing the upper sign reproduces the same equation with the 
exchange $\left(  c_0, c_1 \right) \rightarrow \left( -c_0, - c_1 
\right)$. 
Since $c_{0,1}$ are arbitrary non-zero coefficients, this sign change is 
immaterial. 

In actual fact, one can use the simpler Lagrangian
\be
L_2 \left( u, \dot{u} \right) = \dot{u}^2 +c_0^2 u^4 +c_1^2 + 2 \, c_0 \, 
c_1 u   \label{L2}
\ee
and the associated Hamiltonian 
\be
{\cal H}_2 =  \dot{u}^2 -c_0^2 u^4 -c_1^2 - 2 \, c_0 \, c_1 
u \,.
\ee
Again, setting ${\cal H}_2=0$ reproduces the Riccati 
equation~(\ref{Riccati}).

The equation 
${\cal H}_2=0$ lends itself to a new analogy with point 
particle mechanics;\footnote{The Hamiltonian ${\cal H}_1$ lends itself to 
an analogy with the one-dimensional motion of a particle with mass 
dependent on the position, which is not as compelling, and this is 
the  reason why we switch to the Lagrangian $L_2$ instead of using 
$L_1$ in this mechanical analogy.}  it 
can be seen as the energy conservation equation for 
a particle of unit mass in one-dimensional motion along the $u$-axis in 
the potential energy 
\be
V(u) = - c_0 \, u \left( \frac{c_0}{2} \, u^3 + c_1 \right)
\ee
and with kinetic energy $\dot{u}^2/2$. This energy conservation equation 
is
\be
\frac{\dot{u}^2}{2} +V(u) = E =\frac{ c_1^2}{2}
\ee
for the special value of the total mechanical energy $E=c_1^2/2 > 0$.
 
Consider first the case $c_0 \, c_1>0$: then the potential $V(u) $ has an 
absolute maximum 
\be
V_\mathrm{max}= \frac{ 3 \, c_0^{2/3} c_1^{4/3}}{2^{7/3}} > 0
\ee
at $ u_\mathrm{max}= \mp \left| \frac{c_1}{2\, c_0}\right|^{1/3} <0$ 
(the sign of $u_\mathrm{max}$ depends on the signs of $c_0$ and $ c_1 $).  
The 
motion is always unbounded: there are no turning points if 
$V_\mathrm{max} \leq E$, corresponding to $\frac{c_0}{c_1} \leq 
\frac{4}{\sqrt{27}}$, and there are two turning points otherwise. The same 
conclusion is reached for $c_0 \, c_1 <0 $.

\section{Concluding remarks}
\label{sec:4}

The Riccati equation describes several physical systems, for example the 
vertical speed $v(t)$ of a falling raindrop subject to gravity and 
friction quadratic in the velocity ({\em e.g.}, \cite{myexbook,Rosu}). Let 
$g$ be the constant acceleration of gravity and consider a vertical axis 
pointing downwards, then Newton's second law is \be m \, \frac{dv}{dt} = 
mg-\alpha \, v^2 \,, \ee where $m$ is the mass of the drop and $\alpha $ 
is a friction coefficient. The function $v(t)$ satisfies the Riccati 
equation $\dot{v} +\frac{\alpha}{m} \, v^2 -g=0 $, and one would naively 
think that it is sufficient to write down the Lagrangian for this particle 
to obtain the Lagrangian for the Riccati equation, but including quadratic 
(or, in general, non-linear) friction in the Lagrangian formalism is not 
so easy \cite{Goldstein}. Indeed, the Riccati Lagrangians $L_1$ or $L_2$ 
provided by Eqs.~(\ref{L1}) and~(\ref{L2}) solve this problem of physical 
interest, as well as  that of the many physical systems described by 
Eq.~(\ref{Riccati}), even though the analogous universe is essentially of 
no relevance for physical cosmology. The Friedmann equation describing 
this universe is formally a Riccati equation, but this cosmos is 
unphysical because its parameters must be tuned in order for the analogy 
to hold (the cases studied 
are the only ones for which the Friedmann equation 
{\em in  cosmic time} assumes the Riccati 
form~(\ref{Riccati})). This is 
not an issue here since our goal is not to describe the real universe with 
a Riccati equation (which is usually done by 
rewriting a combination of the Friedmann 
equation and of the acceleration equation   
in conformal time \cite{Faraoni:1999qu,Barrow1993,JantzenUggla}), 
but rather to solve the Helmoltz problem for Eq.~(\ref{Riccati}). The 
explicit Lagrangian and Hamiltonian for the Riccati equation are very 
simple, but one could not guess them without the analogy.

\begin{acknowledgements} 

The author is grateful to a referee for several comments improving the 
presentation of this material. This work is supported, in part, by the 
Natural Sciences \& Engineering Research Council of Canada (grant no. 
2016-03803).

\end{acknowledgements}




\begin{thebibliography}{}

\bibitem{Helmoltz} H. von Helmholtz, ``Ueber die physikalische Bedeutung 
des Princips der kleinsten Wirkung, J. Reine Angewandte Math. {\bf 100}, 
18 (1887).

\bibitem{Douglas} J. Douglas, ``Solution of the inverse problem of the 
calculus of variations'', Trans. Am. Math. Soc. {\bf 50}, 71--128 (1941).

\bibitem{Sarlet} W. Sarlet, ``The Helmholtz conditions revisited. A new 
approach to the inverse problem of Lagrangian dynamics'', J. Phys. A: 
Math. Gen. {\bf 15}, 1503--1517 (1982).

\bibitem{Prince} G.E. Prince, D. King, ``The inverse problem in the 
calculus of variations: nonexistence of Lagrangians'', in {\em 
Differential Geometric Methods on Mechanics and Field Theory: Volume in 
Honour of Willy Sarlet}, F. Cantrijn and B. Langerock editors (Academia 
Press, Gent, Belgium, 2007), pp. 131--140.

\bibitem{Zenkov} D.V. Zenkov (editor), {\em The Inverse Problem of the 
Calculus of Variations: Local and Global Theory} (Atlantis Press, Paris, 
2015).

\bibitem{Carinena} J.F. Cari\~nena, M.F. Ra\~nada, M. Santander, 
``Lagrangian Formalism for nonlinear second-order Riccati Systems: 
one-dimensional Integrability and two-dimensional Superintegrability'', J. 
Math. Phys. {\bf 46}, 062703 (2005).

\bibitem{Musielak} Z.E. Musielak, ``Standard and non-standard Lagrangians 
for  dissipative dynamical systems with variable coefficients'',  J. 
Phys.~A: Math. Theor. {\bf 41}, 055205 (2008).

\bibitem{Cieslinski} J.L. Cie\'sli\'nski, T. Nikiciuk, ``A direct 
approach to the construction of standard and non-standard Lagrangians for 
dissipative-like dynamical systems with variable coefficients'', J. Phys. 
A: Math. Theor. {\bf 43}, 175205 (2010).

\bibitem{Musielak2} Z.E. Musielak, ``General conditions for the existence 
of non-standard Lagrangians for dissipative dynamical systems'', Chaos, 
Sol. Fract. {\bf 42}, 2645-52 (2009).

\bibitem{Wald} R.M. Wald, {\em General Relativity} (Chicago University 
Press, Chicago, 1984).  

\bibitem{Liddle} A. Liddle, {\em An Introduction to Modern Cosmology}  
(Wiley, New York, 2015).

\bibitem{Faraoni:1999qu} V.~Faraoni, ``Solving for the dynamics of the 
universe,'' Am. J. Phys. \textbf{67}, 732 (1999) doi:10.1119/1.19361 
[arXiv:physics/9901006 [physics]].

\bibitem{Ince} E.L. Ince, {\em Ordinary Differential Equations} (Dover, 
New York, 1944), pp. 23--25.

\bibitem{Hille} E. Hille, {\em Lectures on Ordinary Differential 
Equations} (Addison--Wesley, Reading, MA, 1969), pp. 273--288.

\bibitem{Pudasaini} S.P. Pudasaini, M. Krautblatter, ``The 
Landslide Velocity'',  arXiv:2103.10939.

\bibitem{Pudasaini2} S.P. Pudasaini, M. Krautblatter, ``The 
Mechanics of Landslide Mobility with Erosion'', arXiv:2103.14842.

\bibitem{geomagnetic} S. Kov\'a\u{c}ikov\'a, J. Pek, ``Generalized 
Riccati equations for 1-D magnetotelluric impedances over anisotropic 
conductors Part I: plane wave field model'',  Earth, Planets \& Space 
{\bf 54}, 473--482 (2002).

\bibitem{DenmarkStrait} R. K\"ase, ``A Riccati model for 
Denmark Strait  overflow variability: Thermohaline circulation variability 
in the subpolar  North Atlantic'',  Geophys. Res. Lett. {\bf 33}, 21 
(2006).

\bibitem{Schuch} D. Schuch, ``Relations between nonlinear Riccati 
equations and other equations in fundamental physics'', J. Phys.: Conf. 
Ser. {\bf 538}, 012019 (2014).

\bibitem{Schuch2} D. Schuch, {\em Quantum Theory from a Nonlinear 
Perspective} (Springer, New York, 2018).

\bibitem{Rosu} M. Novakowski, H.C. Rosu, ``Newton's laws of motion in 
form of Riccati equation'', Phys. Rev. E {\bf 65}, 047602 (2002).

\bibitem{myexbook} V. Faraoni, {\it Exercises in Environmental Physics} 
(Springer, New York, 2006).

\bibitem{deRitis1} S. Capozziello, R. de Ritis, ``Relation between the 
potential and nonminimal coupling in inflationary cosmology'', Phys. Lett. 
A {\bf 177}, 1 (1993).

\bibitem{deRitis2} S. Capozziello, R. de Ritis, ``N\"other's symmetries 
and exact solutions in flat non-minimally coupled cosmological models'', 
Class. Quantum Grav. {\bf 11}, 107 (1994).

\bibitem{deRitis3} M. Demianski, R. de Ritis, G. Platania, C. Rubano, P. 
Scudellaro, C. Stornaiolo, ``Scalar field, nonminimal coupling, and 
cosmology'', Phys. Rev. D {\bf 44}, 3136 (1991).

\bibitem{SanyalModak} A.K. Sanyal, B. Modak, ``Is Noether Symmetric 
Approach Consistent With Dynamical Equation In Nonminimal Scalar-Tensor 
Theories?'', Class. Quantum Grav. {\bf 18}, 3767 (2001).

\bibitem{mybook} V. Faraoni, {\em Cosmology in Scalar-Tensor Gravity} 
(Kluwer Academic, Dordrecht, 2004).

\bibitem{Caldwell:1999ew} R.R.~Caldwell, ``A Phantom menace?,'' Phys. 
Lett. B \textbf{545}, 23-29 (2002) doi:10.1016/S0370-2693(02)02589-3 
[arXiv:astro-ph/9908168 [astro-ph]].

\bibitem{Caldwell:2003vq} R.R.~Caldwell, M.~Kamionkowski,  
N.N.~Weinberg, ``Phantom energy and cosmic doomsday,'' Phys. Rev. Lett. 
\textbf{91}, 071301 (2003) doi:10.1103/PhysRevLett.91.071301 
[arXiv:astro-ph/0302506 [astro-ph]].

\bibitem{Barrow1993} J.D. Barrow, ``Relativistic cosmology and the 
regularization of orbits'', {\em The Observatory} {\bf 113}, 210 (1993).

\bibitem{JantzenUggla} R.T. Jantzen, C. Uggla, ``Structure of the 
generalized Friedmann problem'', Gen. Relativ. Gravit. {\bf 24}, 59 
(1992).

\bibitem{Euler} L. Euler, ``De motu rectilineo trium corporum se mutuo 
attrahentium.'', {\em Novi Comm. Acad. Sci. Petrop.} {\bf 11}, 144 (1765).

\bibitem{Bohlin1911} K. Bohlin, ``Note sur le probl\'eme des deux corps 
et sur une int\'egration nouvelle dans le probl\'eme des trois corps'', 
{\em Bull. Astron.} {\bf 28}, 113 (1911).

\bibitem{Sundman1912} K. Sundman, ``Memoire sur le probleme de trois 
corps'', {\em Acta Math.} {\bf 36}, 105 (1912).

\bibitem{LeviCivita1920} T. Levi-Civita, ``Sur la r\'egularisation du 
probl\`eme des trois corps'', {\em Acta Math.} {\bf 42}, 99--144 
(1920), doi:10.1007/BF02404404

\bibitem{Bond1985} V.R. Bond, ``A transformation of the two-body 
problem'', {\em Celestial Mechanics} {\bf 35}, 1--7 (1985), 
https://doi.org/10.1007/BF01229108

\bibitem{HeggieHut2003} D. Heggie, P. Hut, {\em The Gravitational 
Million-Body Problem} (Cambridge University Press, Cambridge, 2003), pp. 
143--149 (2003).

\bibitem{Goldstein} H. Goldstein, {\it Classical Mechanics} 
(Addison--Wesley, Reading, Massachusetts, 1980).

\end{thebibliography}


\end{document}